\documentclass[acmtog,review=false,nonacm]{acmart}

\usepackage{booktabs} 
\usepackage{soul}
\usepackage{xcolor,colortbl}
\usepackage{wrapfig}

\usepackage[ruled]{algorithm2e}

\SetAlFnt{\small}
\SetAlCapFnt{\small}
\SetAlCapNameFnt{\small}
\SetAlCapHSkip{0pt}

\acmJournal{TOG}

\def\figurePath{figures/}

\def\myfigure#1#2{\begin{figure}[t]\centering\includegraphics*[width = \linewidth]{\figurePath#1}\vspace{4pt}\caption{#2 }\label{fig:#1}\vspace{-3pt}\end{figure}}
\def\mycfigure#1#2{\begin{figure*}[t]\centering\includegraphics*[clip, width = \linewidth]{\figurePath#1}\vspace{4pt}\caption{#2 }\label{fig:#1}\vspace{-3pt}\end{figure*}}

\def\mysection#1#2{\section{#1}\label{sec:#2}}

\def\mysubsection#1#2{\subsection{#1}\label{sec:#2}}

\newcommand{\eg}{e.g.,\ }
\newcommand{\etal}{ \textit{et al}. }
\newcommand{\refSec}[1]{Section~\ref{sec:#1}}
\newcommand{\refFig}[1]{Figure~\ref{fig:#1}}

\newcommand{\refTbl}[1]{Table~\ref{tbl:#1}}

\newcommand{\degree}{^\circ}

\definecolor{grn}{rgb}{0.8,1.0,0.8}
\definecolor{rd}{rgb}{1.0,0.8,0.8}
\definecolor{ylw}{rgb}{1.0,1.0,0.8}

\author{Taimoor Tariq}
\affiliation{%
\department{Perception, Display and Fabrication Group}
\institution{Università della Svizzera italiana, Switzerland}}
\email{tariqt@usi.ch}

\author{Cara Tursun}
\affiliation{%
\institution{Università della Svizzera italiana, Switzerland}}

\author{Piotr Didyk}
\affiliation{%
\institution{Università della Svizzera italiana, Switzerland}}

\begin{document}

\title{Noise-based Enhancement for Foveated Rendering}

\begin{teaserfigure}
\includegraphics[width=\textwidth]{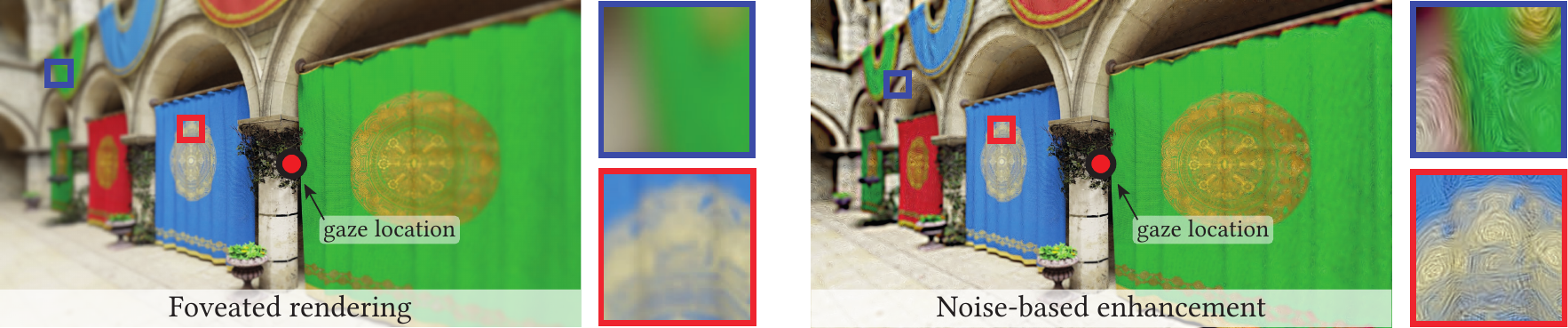}
\caption{Our method enhances the output of standard foveated rendering (left) by adding carefully tuned procedural noise (right). The enhancement replaces accurate rendering of spatial details, which are detectable but not resolvable by the human visual system, and allows applying more aggressive foveation during the rendering process.}
\label{fig:teaser}
\end{teaserfigure}

\begin{abstract}
Human visual sensitivity to spatial details declines towards the periphery. Novel image synthesis techniques, so-called foveated rendering, exploit this observation and reduce the spatial resolution of synthesized images for the periphery, avoiding the synthesis of high-spatial-frequency details that are costly to generate but not perceived by a viewer. However, contemporary techniques do not make a clear distinction between the range of spatial frequencies that must be reproduced and those that can be omitted. For a given eccentricity, there is a range of frequencies that are detectable but not resolvable. While the accurate reproduction of these frequencies is not required, an observer can detect their absence if completely omitted. We use this observation to improve the performance of existing foveated rendering techniques. We demonstrate that this specific range of frequencies can be efficiently replaced with procedural noise whose parameters are carefully tuned to image content and human perception. Consequently, these frequencies do not have to be synthesized during rendering, allowing more aggressive foveation, and they can be replaced by noise generated in a less expensive post-processing step, leading to improved performance of the rendering system. Our main contribution is a perceptually-inspired technique for deriving the parameters of the noise required for the enhancement and its calibration. The method operates on rendering output and runs at rates exceeding 200\,FPS at 4K resolution, making it suitable for integration with real-time foveated rendering systems for VR and AR devices. We validate our results and compare them to the existing contrast enhancement technique in user experiments.
\end{abstract}

%
%

%
%


\maketitle

\mysection{Introduction}{introduction}
The quality of rendered images directly impacts user experience, immersion, and comfort. While today's computer graphics techniques offer endless opportunities to improve image quality, either by more accurate lighting simulation or modeling of the virtual world, additional constraints, such as frame rate or power efficiency, can significantly limit image quality. Rendering efficiency is critical in the context of new virtual and augmented reality headsets that require rendering systems to operate at high spatial and temporal resolutions. Therefore, it is essential to save computational resources whenever possible, ideally matching the quality of the images with the capabilities of human perception.

Significant computational savings can be achieved when the rendering quality is guided by the estimation of viewers' gaze location provided by an eye-tracking device. So-called foveated rendering techniques \cite{monhanto2021} reduce the rendering quality according to the decay in the sensitivity of the human visual system (HVS) to image distortions in the periphery. The most common approach is to reduce the shading rate or spatial resolution for image regions distant from the gaze location. This process aims to omit the costly synthesis of high-spatial frequency details irrelevant for human perception. The benefits of foveated rendering are already being exploited by the latest VR displays such as the Sony Play-Station VR2\footnote{\url{https://blog.playstation.com/2022/01/04/playstation-vr2-and-playstation-vr2-sense-controller-the-next-generation-of-vr-gaming-on-ps5/}} and Varjo VR-3\footnote{\url{https://developer.varjo.com/docs/native/foveated-rendering-api}}.

\myfigure{thibos_aliasing}{Subjective appearance of gratings located at $20\degree$ retinal eccentricity in the nasal visual field. Redrawn from Thibos\etal\shortcite{thibos1998}.}

Unfortunately, there is no clear distinction between the spatial frequencies relevant to a human observer and the spatial frequencies that can be excluded from the rendering procedure. In fact, there is a range of spatial frequencies that are detectable by the HVS but not resolvable, i.e., the human sensitivity to its spatial localization and orientation is low \cite{thibos1987}. This means that even though these frequencies do not need to be accurately synthesized, they cannot be omitted in the process of foveated rendering since this would lead to visible quality degradation (Figure~\ref{fig:thibos_aliasing}). As a result, in order to provide images perceptually equivalent to full-resolution rendering, foveated rendering techniques must synthesize this range of frequencies.

Inspired by the existence of spatial frequencies that are detectable but not resolvable, we argue that a direct rendering of this range of spatial details is not necessary and can be replaced by an inexpensive procedural noise synthesis. Consequently, we propose a foveated rendering procedure where we first use foveated rendering to synthesize the image content that falls into the spatial frequency range which is both detectable and resolvable by a human observer. Next, we use the synthesized information to estimate the parameters of the noise, which we synthesize and combine with the previously rendered content in a post-processing enhancement step. Figure~\ref{fig:human_perception} shows the human visual acuity regions, and how they are addressed by different rendering approaches. The main contribution of our paper lies in the derivation of the method for estimating the noise parameters, based on the underlying content. Our procedure is based on the perceptual literature and consists of several simple image processing steps, making it efficient to implement and temporally coherent. The method is calibrated by a series of perceptual experiments to fine-tune the range of spatial frequencies added with the noise and their amplitudes. In additional evaluation experiments, we demonstrate that our noise-based enhancement step can successfully improve the fidelity of aggressive foveated rendering. We also compare our method to the recent work on contrast enhancement for foveated rendering \cite{patney2016}.


\mysection{Background and Related Work}{related_work}
In this section, we introduce the previous studies on the characteristics of the peripheral vision, which form the perceptual basis of our work (\refSec{retina}). We briefly discuss the existing applications of foveated rendering (\refSec{foveated_rendering}) and introduce works on metamerism (\refSec{metamerism}).

\myfigure{human_perception}{Characterization of spatial acuity for peripheral vision. Resolution acuity, detection acuity, and the region of spatial aliasing are shown as a function of retinal eccentricity \cite{thibos1996}.}

\mysubsection{Peripheral Vision}{retina}
While observing our environment if we fixate at a point with our eyes, we notice that we cannot see everything at high acuity in our visual field \cite{strasburger2011,rosenholtz2016}. This is due to a decreasing limit on the visual acuity as the distance from the point of fixation increases \cite{aubert1857, hering1899}. Our eyes feature a central high acuity region called \emph{fovea} that is approximately $2\degree$ wide and \emph{peripheral vision} that covers whole hemisphere outside fovea with low acuity \cite{lettvin1976}. The decreasing peripheral visual acuity was systematically measured using square-wave spatial contrast gratings at different retinal eccentricities \cite{wertheim1894} and those measurements were initially related to sampling rate determined by the density of retinal rod and cone type of photoreceptor cells \cite{fick1898,osterberg1935,hirsch1989,curcio1990a,levi1986}. More recent studies show that the contrast sensitivity in the peripheral vision can be as high as those measured at the fovea, provided that the stimulus size is magnified depending on an estimate of ganglion cell density in the periphery \cite{rovamo1978,rovamo1979, virsu1987,legge1987,rossi2010}. These studies on the retinal anatomy have led to models of peripheral sensitivity for computing visibility across the visual field \cite{watson2016,schutt2017,watson2018,haun2021}.

The neural resolution of the photoreceptor cells is only one of the factors defining the spatial frequency limits of the human eye. The other factor is the optical system mainly represented by the cornea and the lens. The neural resolution limits were measured in isolation from optical aberrations using an interferometer for both fovea and periphery \cite{williams1985a, williams1985b,thibos1985}. Although both of the optical quality and neural resolution decline with retinal eccentricity, the decline in neural resolution is more significant than the optical limit \cite{green1970, millodot1975, anderson1991}. As a result, the photoreceptors in peripheral vision may get stimulated by spatial frequencies beyond their resolution limit, which manifests itself as aliasing \cite{thibos1987b}. Beyond resolvable spatial frequency levels, the presence of the stimuli in the peripheral vision can still be detected; however, low-level details are not resolvable (\eg orientation of isolated gratings) \cite{anderson2002}. In experiments with filtered images, this may lead to participants reporting a subjective visual change in the image without being able to define what exactly changed \cite{sere2000}. Thibos\etal\shortcite{thibos1987, thibos1998} studied the perception of signals beyond neural sampling limit and identified the range of spatial frequencies where visual stimuli are detectable but not resolvable (\refFig{human_perception}). 






\mysubsection{Foveated Rendering}{foveated_rendering}
Foveated rendering techniques reduce the rendering quality according to the loss of human visual sensitivity in the periphery. Earlier applications involved image-video compression, spatially adaptive ray tracing and tuning parameters such as the resolution and bit-depth \cite{tong1984,browder1988,levoy1990,glenn1994,tsumura1996,kortum1996}. More recently, Guenter\etal\shortcite{guenter2012} demonstrated that eccentricity-dependent shading rate reduction can lead to computational savings without significant quality degradation. Stengel\etal\shortcite{stengel2016} made use of additional information from the geometry pass in rendering, such as depth, normal, and texture properties, to derive the local information on silhouettes, object saliency, and specular highlights from the last frame to reduce the shading rate. Patney\etal\shortcite{patney2016} introduced a cost effective method based on contrast enhancement of attenuated spatial frequencies that significantly reduces the visibility of peripheral blur in foveated rendering. Swafford\etal\shortcite{swafford2016} extended the HDR-VDP2 \cite{mantiuk2011} quality metric for peripheral vision and introduced a method for tuning ambient occlusion, tessellation and ray-casting quality. Meng\etal\shortcite{meng2018} used kernel log-polar mapping for an efficient GPU-based implementation of foveated rendering. Tursun\etal\shortcite{tursun2019} modeled the influence of underlying content on the visibility and proposed a model that takes both the retinal eccentricity and the content into account. Kaplanyan\etal\shortcite{kaplanyan2019} showed an application of  neural networks on foveated image reconstruction with very sparse sampling but with a significant computational overhead similar to other neural reconstruction techniques. For a comprehensive survey of foveated rendering methods, please refer to the work of Mohanto\etal\shortcite{monhanto2021}.



\mysubsection{Visual Metamerism}{metamerism}
\emph{Visual metamerism} refers to the phenomena of physically different images being perceptually indistinguishable. The concept has recently attracted research interest, aimed at finding perceptually equivalent stimuli that are computationally simpler to reconstruct in the context of gaze-contingent applications. Initial studies on the neural representation of vision hypothesized that the role of early vision is to have an efficient representation by removing statistical redundancy in natural images \cite{barlow1961,simoncelli2001}. Portilla and Simoncelli\shortcite{portilla2000} showed that a statistical model based on wavelet representations can be used to synthesize visually plausible natural and artificial textures. Inspired by Portilla and Simoncelli, there have been more recent studies aimed towards synthesizing visual metamers using deep convolutional neural network representations \cite{freeman2011,deza2018}. Very recently, Walton\etal\cite{walton2021} exploited spatial pooling in human peripheral vision and proposed a technique for foveated graphics to synthesize visual metamers. The technique requires the full-resolution image statistics; therefore, its application is limited to tasks where the reference image is available, such as foveated compression, but it is not applicable to the real-time foveated rendering pipeline.

A comparison of our noise-based enhancement method to the most relevant works from the state-of-the-art is shown in \refTbl{comparison}. For foveated rendering, it is essential for a method to avoid a preliminary full-resolution rendering pass to compute the input. Among the compared methods, only Walton\etal\shortcite{walton2021} requires the full-resolution input to compute the image statistics for the output; therefore, it is more appropriate for compression tasks rather than foveated rendering. In addition, a desirable property motivated by the work of Thibos\etal\shortcite{thibos1987,thibos1987b,thibos1996} is enhancing the output by synthesizing novel image details at \emph{detectable \& not resolvable band} (\refFig{human_perception}) that do not already exist in the input image. Among the compared methods, Kaplanyan\etal\shortcite{kaplanyan2019} uses neural reconstruction trained on natural and synthesized images for this task while Walton\etal\shortcite{walton2021} captures the required statistics from the full-resolution reference input. Although Patney\etal\shortcite{patney2016} enhance the input image contrast, they do not add novel spatial details to their output. Regarding running times, due to the computational efficiency of the simple contrast enhancement, Patney\etal\shortcite{patney2016} is the fastest among compared methods. Our method has the second place with our computationally efficient real-time GPU implementation. Walton\etal\shortcite{walton2021} report real-time running times for low resolutions, their application becomes challenging at high resolutions due to costly image analysis step. Kaplanyan\etal\shortcite{kaplanyan2019} is the most costly method among all compared because their network inference requires 4 GPUs to process a $ 1920 \times 1080 $ input at interactive rates.


%



\begin{table}[htb]
    \centering
	\caption{Comparison of our method with relevant works from the literature.}
	\label{tbl:comparison}
	\begin{tabular}{cp{35pt}p{35pt}p{35pt}}
		\toprule
		Method & Requires full-res input & Spatial-Frequency synthesis & Running time \\
		\midrule
		Patney\etal\shortcite{patney2016} & \cellcolor{grn}No & \cellcolor{rd}No & \cellcolor{grn}Fastest \\
		\hline
		Kaplanyan\etal\shortcite{kaplanyan2019} & \cellcolor{grn}No & \cellcolor{grn}Yes & \cellcolor{rd}Slow \\
		\hline
		Walton\etal\shortcite{walton2021} & \cellcolor{rd}Yes & \cellcolor{grn}Yes & \cellcolor{ylw}Moderate \\
		\hline
		Ours & \cellcolor{grn}No & \cellcolor{grn}Yes & \cellcolor{grn}Fast \\
		\bottomrule
	\end{tabular}
\end{table}
\mycfigure{pipeline}{The figure demonstrates our noise-based enhancement for foveated rendering. The input is a foveated image which is processed to estimate parameters for Gabor noise, i.e., orientation, frequency, and amplitude. Resulting Gabor kernels are convolved with random impulses to synthesize procedural noise. Next, the noise is added to the contrast enhanced foveated image.}

\section{Method Overview}
Our method enhances the output of adaptive-resolution foveated rendering techniques \cite{monhanto2021} by synthesizing novel image details whose spatial frequencies fall into the range of detectable but not resolvable frequencies by the HVS (Figure~\ref{fig:human_perception}). We show that these details can be synthesized using carefully controlled procedural noise. While several types of noise can be considered, we rely on sparse Gabor convolution which provides good frequency control and computationally efficient for real-time performance \cite{lagae2009,lagae2010}. The main contribution of this work is the perceptually inspired method for estimating the parameters of the noise based on the underlying content. Our method relies solely on the information available in the foveated image and it does not require a full-resolution rendering pass. This makes it suitable for practical real-time foveated rendering applications.

The overview of our noise-based enhancement is depicted in Figure~\ref{fig:pipeline}. The input to our technique is a single frame from a foveated rendering sequence. We assume that the rendering does not contain significant spatial or temporal aliasing already. Consequently, following the studies of Albert\etal\shortcite{albert2017}, Hoffman\etal \shortcite{hoffman2018} and the previous work on foveated rendering \cite{patney2016,tursun2019}, we assume that the resolution reduction due to foveation can be modeled with a spatially-varying Gaussian filter. We denote the standard deviation of Gaussian filter as $\sigma(\mathbf{x})$, where $\mathbf{x} = [x_h\, x_v]$ is the vector of normalized horizontal and vertical image coordinates such that $ x_h, x_v \in [0,1] $. Given the input foveated image and $\sigma(\mathbf{x})$, our technique estimates the local noise parameters: amplitude, spatial frequency, and orientation. The parameters are later used to generate the additive noise for the input. In a separate step, we apply the contrast enhancement \cite{patney2016} to the foveated image. The technique effectively enhances the spatial frequencies that have been attenuated, but not removed by the foveated rendering. In the final step, we add the generated noise to the contrast-enhanced version of the input image to introduce novel spatial details in the aliasing band.

Below, we describe the method for noise-based enhancement (Section~\ref{sec:noise_based_enhancement}) and its calibration to the data collected from perceptual experiments (Section~\ref{sec:calibration}). For details of the procedural noise generation and contrast enhancement methods that we use, please refer to the original papers \cite{lagae2009,patney2016}.

\section{Noise-Based Enhancement}
\label{sec:noise_based_enhancement}
In our work, we use sparse Gabor convolution to generate the noise \cite{lagae2009}. The method convolves Gabor kernels with sparse, random impulses (Figure~\ref{fig:pipeline}). This is equivalent to placing Gabor patches at random locations across the image. The characteristics of the noise are controlled by the properties of individual Gabor kernels, i.e., amplitude, spatial frequency, and orientation. In this section, we describe our technique for estimating these parameters from a foveated rendering input. For a given location of an impulse $\mathbf{x}$ in the image, our method estimates the spatial frequency, $f(\mathbf{x})$, amplitude, $K(\mathbf{x})$, and orientation, $\omega(\mathbf{x})$ of the Gabor kernel convolved with this impulse. 

It is critical that these parameters are not estimated using a full-resolution image as that would require a full-resolution rendering pass and hinder any potential computational gains from foveated rendering by making it as slow as standard full-resolution rendering. Instead, we assume that there is a sufficient correlation between the low and high-spatial frequency components of an image that allows us to derive the noise parameters based on the foveated image. This assumption is motivated by the fact that the power spectrum of natural images follows a power-law \cite{ruderman1994,hyvarinen2009}. In addition, such an assumption has already been used in the context of foveated rendering to estimate missing luminance contrast \cite{tursun2019}. While it does not hold in all cases (Section~\ref{sec:limitations}), we demonstrate its successful application in our scenario.

\subsection{Frequency}
\label{sec:frequency}
Our method generates noise that is detectable but not resolvable by the HVS (Figure~\ref{fig:human_perception}). We follow the study of Thibos\etal\shortcite{thibos1996} which provides the range of spatial frequencies with these properties as a function of eccentricity. We denote the upper limit of the range by $T_H(e_\mathbf{x})$ and the lower limit by $T_L(e_\mathbf{x})$, where $e_\mathbf{x}$ is the retinal eccentricity at location $\mathbf{x}$ in the input image, i.e., the distance from the gaze location measured in visual degrees. While the original work provides only a few measured data points of the limits (Section~\ref{sec:thibos_measurements}), we linearly interpolate the measurements to obtain $T_H(e_\mathbf{x})$ and $T_L(e_\mathbf{x})$ for a wider range of eccentricities. The result of the interpolation is shown in Figure~\ref{fig:human_perception}.

To respect the limits defined by Thibos\etal\shortcite{thibos1996}, we define the bounds of the spatial frequency of the noise, $f(\mathbf{x})$, as $T_H(e_\mathbf{x})$ and $T_L(e_\mathbf{x})$. Additionally, $f(\mathbf{x})$ should also be bounded by the range of frequencies that are absent from or attenuated by foveated rendering. This additional lower bound on $f(\mathbf{x})$ can be computed from the spatial frequency cut-off point depending on the rendering resolution of foveated rendering or in our case, from the standard deviation of Gaussian kernel, $\sigma(\mathbf{x})$, as we use this low-pass filter to simulate the adaptive-resolution effects of foveation. We use the fact that the amplitude attenuation of a Gaussian blur is also a Gaussian in the frequency domain with standard deviation $\sigma_f = \frac{1}{2\pi\sigma}$, where $\sigma$ is the standard deviation of the Gaussian in the spatial domain. The Gaussian filter has an infinite-band impulse response and it does not have a sharp frequency cut-off. In our work, we define the cut-off frequency at $3\sigma_f$, where the filter response is approximately 1\% of the peak. Consequently, we compute the lower limit on the noise frequency as the maximum of the lower bound from Thibos\etal and the cut-off frequency:
\begin{equation}
F_L(\mathbf{x}) = \max\left(T_L(e_\mathbf{x}), \frac{3}{2\pi\sigma(\mathbf{x})}\right).
\end{equation}
It is common to define the cut-off frequency at the 50\% of the peak filter response in signal processing applications. We make this conservative selection of the cut-off point at $3\sigma_f$ to avoid adding noise to the resolvable frequency range because it would be perceived as undesired distortions to the image by observers. Apart from the frequency spectrum of the foveated image, the noise added to the image must also respect the physical limitations of the display. Consequently, we define the upper limit on the noise frequency to be the minimum of the upper bound from Thibos\etal and the limit of the display:
\begin{equation}
F_H(\mathbf{x}) = \min\left(T_H(e_\mathbf{x}), f_d\right),
\end{equation}
where $f_d$ is the highest spatial frequency the display can reproduce. According to the Nyquist–Shannon sampling theorem, this limit is equal to 0.5 cycles per pixel. Note that for brevity in our derivation, we omit conversion between different units of spatial frequency, which are either cycles per visual degree or cycles per pixel. The equations hold as long as the units are same, and the conversion is performed based on the display pixel size and the viewing distance of an observer. 

Given the theoretical frequency limits on the noise ($F_L(\mathbf{x})$ and $F_H(\mathbf{x})$), we model the local noise frequency $f(\mathbf{x})$ as a random variable $\chi_\mathbf{x}(f)$. In the early vision, the bandwidth of cortical simple cells is symmetrical in log spatial frequency domain \cite{devalois1982}. In order to establish a closer resemblance to this bandwidth symmetry, we define the distribution of $\chi_\mathbf{x}(f)$ as log-normal, with the mean, $\mu_n(\mathbf{x})$, at the midpoint of $F_L(\mathbf{x})$ and $F_H(\mathbf{x})$ in log domain and standard deviation, $\sigma_n(\mathbf{x})$, as the quarter of this range such that the range between $F_H(\mathbf{x})$ and $F_L(\mathbf{x})$ is spanned by $\mu_n(\mathbf{x})\pm2\sigma_n(\mathbf{x})$ in log frequency domain. In order to fine tune the bandwidth around $\mu_n(\mathbf{x})$, we additionally define a scaling factor $s_f$ as a free parameter of our model for adjusting $\sigma_n(\mathbf{x})$:
\begin{equation}
  \mu_n(\mathbf{x}) = 0.5 \cdot (\ln F_L(\mathbf{x}) + \ln F_H(\mathbf{x})),
\end{equation}
\begin{equation}
  \sigma_n(\mathbf{x}) = 0.5 \cdot s_f \cdot (\mu_n(\mathbf{x}) - \ln F_L(\mathbf{x})).
\end{equation}

\begin{figure}
\includegraphics[width=\linewidth]{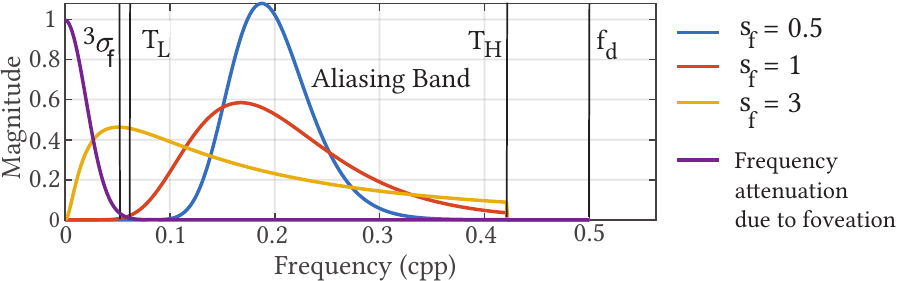}
\caption{Examples of different distributions of random variable $\chi_\mathbf{x}(f)$, for Gabor kernels, with all considered limits.}
\label{fig:freq_explanation}
\end{figure}

Using these mean and standard deviation parameters, we formally define the distribution of $\chi_\mathbf{x}(f)$ as:
\begin{equation}
    \chi_\mathbf{x}(f) \sim e^{\mu_n(\mathbf{x}) + \sigma_n(\mathbf{x}) Z},
\end{equation}
where $Z$ is the normal distribution $\mathcal{N}(0,\,1)$. Figure~\ref{fig:freq_explanation} shows all the limits considered when estimating the frequency distribution, $\chi_\mathbf{x}(f)$, as well as three examples of the distributions for different values of $s_f$. 

We calibrate bandwidth scaling factor, $s_f$, using the participants' responses from a subjective experiment (Section~\ref{sec:calibration}). $s_f$ is a scalar that is independent of location, eccentricity and underlying content as these factors are already considered by $F_L(\mathbf{x})$ and $F_H(\mathbf{x})$ in our model. Since the distribution of $\chi_\mathbf{x}(f)$ has infinite falloff, we additionally truncate it to range $(0,F_H(\mathbf{x}))$. While generating the noise, for every location of an impulse, $\mathbf{x}$, we compute the corresponding spatial frequency $f(\mathbf{x})$ of the Gabor patch by randomly sampling from the truncated log-normal distribution of  $\chi_\mathbf{x}(f)$.

\subsection{Amplitude}
While estimating the local noise amplitude, $K(\mathbf{x})$, we aim to synthesize novel details, which resemble those in full-resolution rendering. The challenge comes from the fact that our noise-based enhancement generates spatial frequencies that are not present in the foveated input. Inspired by previous work \cite{tursun2019} and natural image statistics literature (Section~\ref{sec:related_work}), we assume that there is a sufficient correlation between low and high frequency components of an image to infer the amplitude of higher frequencies from lower frequency information. More precisely, we propose to estimate the noise amplitude $K(\mathbf{x})$ based on the amplitude that corresponds to the highest frequency component generated by the foveated rendering around location $\mathbf{x}$.  

To efficiently analyze the amplitude of different frequency bands in the input foveated image, we compute its Laplacian pyramid decomposition, $L$, where $L_l(\mathbf{x})$ is the value at the level $l$ at the location $\mathbf{x}$. We treat each level $l$ as a frequency band with cut-off frequency $2^{-l}$ cycles per pixel, and the central frequency 
\begin{equation}
\label{eq:central}
f^l = 2^{-(l + 0.5)},    
\end{equation}
which is the mean of the cut-off frequencies of two consecutive bands in log domain. It is important to remember that while the Laplacian pyramid does not provide strict separation between different spatial frequency bands \cite{Peli1990ContrastIC}, it can be efficiently implemented on a GPU using MIP maps. 

In order to estimate the amplitude of the noise at location $\mathbf{x}$, we first calculate the highest spatial frequency present in the foveated image. As discussed in Section~\ref{sec:frequency}, the frequency attenuation due to foveation can be modeled in the frequency domain using a Gaussian function with $\sigma_f(\mathbf{x}) = \frac{1}{2\pi\sigma(\mathbf{x})}$. Given the Gaussian function modeling the frequency attenuation and the attenuation factor $a$, we can solve for the frequency value that was attenuated with this factor. The solution gives the frequency value:
\begin{equation}
    f_{c}(\mathbf{x}) = \sqrt{-2 \cdot \sigma_f(\mathbf{x}) \cdot \ln(a)} = 
    \sqrt{\frac{-\ln(a)}{\pi\sigma(\mathbf{x})}}.
\end{equation}
In other words, the above formula estimates the cut-off frequency due to foveation at location $\mathbf{x}$ for a given attenuation cut-off, $a$, beyond which we consider the spatial frequencies to be absent or unreliable. Furthermore, by combining the above equation with Equation~\ref{eq:central}, we can compute the Laplacian level $l_a(\mathbf{x})$ that represents the highest reliable spatial frequencies:
\begin{equation}
\label{eq:level}
l_a(\mathbf{x}) = \log_2\left(\sqrt{\frac{-\pi\sigma(\mathbf{x})}{\ln(a)}}\right) - 0.5.
\end{equation}
We use this equation to infer the amplitude of the noise. We assume that the amplitude of the spatial frequencies omitted in the foveated rendering is directly related to the amplitude of the frequencies that are present in the foveated image. Consequently, we estimate the noise amplitude $K(\mathbf{x})$ as: 
\begin{equation}
\label{eq:amplitude}
    K(\mathbf{x}) = s_k \cdot |L_{l}(\mathbf{x})| \quad \text{with} \quad l = l_a(\mathbf{x}),
\end{equation}
where $s_k$ is a constant parameter that is calibrated in the subjective experiment (Section~\ref{sec:calibration}). There is a trade-off in setting the attenuation cut-off, $a$. Higher values make the estimation of $K(\mathbf{x})$ based on the lower frequencies. These are more reliable, but they may correlate less with the amplitude of high frequencies we seek to estimate. While this correlation is potentially better for low values of $a$, the content of these frequencies might be less reliable. For all our experiments, we set $a = 0.25$, which provides good trade-off. It also is important to mention that the value $l_a(\mathbf{x})$ is not guaranteed to be an integer. For non-integer values, to compute Equation~\ref{eq:level}, we perform a linear interpolation between two Laplcian pyramid levels in the log domain.

The above derivation does not account for the fact that adding noise may result in clipping the image values. This is critical for our technique which relies on precise control of noise frequency spectrum and any untreated clipping may significantly alter it. To prevent changes to the noise frequency spectrum, we propose to further attenuate the estimated noise amplitude according to the minimum and maximum image values in a small neighborhood. More precisely, for a give location $\mathbf{x}$, we compute a minimum, $N_{\min}(\mathbf{x}$), and maximum, $N_{\max}(\mathbf{x}$), image values in a small image neighborhood. Given these values, we can estimate the remaining range of image values available for the noise as $\min(1-N_{\max}(\mathbf{x}), N_{\min}(\mathbf{x}))$. Using this range, we scale previously estimated amplitude (Equation~\ref{eq:amplitude}) to respect this range. 

Efficient implementation of estimating minimum and maximum image values in a neighborhood is critical for the real-time performance of our technique. Therefore, we build two additional image pyramids of the foveated image to which we add the noise. One pyramid contains local minimum values of all the image pixels located below, while the other pyramid contains corresponding maximum values. Such pyramids can be efficiently constructed subsequently for each level by using the minimum and maximum values of 2$\times$2 neighborhood from the previous level. Resulting pyramids at level $l$ encode minimum and maximum from $2^l\times2^l$-pixel neighborhood. It is important to note that such a pyramid decomposition is not equivalent to computing exact minimum and maximum values directly at the original resolution, but it serves as computationally efficient approximation of it. In all our results, we use third pyramid levels, which provide a neighborhood size similar to the spatial support of the Gabor patches used for generating the noise.

\subsection{Orientation}
Similar to the amplitude estimation, we seek an orientation estimate, $\omega(\mathbf{x)}$, that closely follows underlying content. Consequently, we compute it from the lower frequency component of the foveated image. To this end, we use the Sobel operator and convolve the foveated image with horizontal and vertical Sobel filters, which provide corresponding image gradients. Then we use them for estimating the local gradient direction. To ensure spatial and temporal smoothness of the gradients, we apply the orientation estimation to the third level of a Gaussian pyramid of the input foveated image and interpolate the results to the original resolution using bicubic interpolation. The Gaussian pyramid does not introduce any additional computational cost since it is an intermediate step when building the Laplacian pyramid for amplitude estimation and we reuse the result of that computation.

\subsection{Noise Generation}
\label{sec:noise_generation}
The noise generation is based on the work of Lagae\etal\shortcite{lagae2009}. The authors propose a technique where Gabor-based noise is generated by convolving a random set of impulses with Gabor kernels. To allow efficient implementation, the impulses are generated in cells that divide the image into a regular grid. While the size of each cell corresponds to the width of the spatial envelope of the Gabor, the number of impulses in each cell can be adjusted to control the trade-off between the quality of the noise and the computational efficiency of the method. The density of impulses is expressed in their number per kernel, i.e., the area of the truncated Gabor kernel. The authors report that 25-50 impulses provide a good quality noise, but their analysis suggests that even lower numbers provide satisfactory results. 

Our generation of noise follows the above procedure closely. We chose the size of our Gabor kernels to accommodate the lowest spatial frequency we aim to generate, which resulted in setting the width of the Gabor patch in the frequency domain to be 0.06 cycles per pixel. The size of the Gabor patch also determined the size of each cell. For all our experiments, we used 64 impulses per kernel unless mentioned otherwise. We use the same random number generator as the original implementation for generating impulses. Before generating the noise, each impulse at location $\mathbf{x}$, is assigned a Gabor kernel parameters: frequency, $f(\mathbf{x})$, amplitude $K(\mathbf{x})$, and orientation $\omega(\mathbf{x})$, estimated using our method described above. To avoid temporal instabilities, the locations of impulses throughout the entire input image sequence are the same. This is achieved by using a constant seed value for the random number generator. Constant positions of impulses ensure that the Gabor patches do not change their position. Only their parameters change smoothly according to the underlying motion in the content.

\subsection{Implementation}
We implemented two versions of our enhancement method. The prototype version was implemented in MATLAB, and the real-time counterpart was implemented in OpenGL. As an input, both implementations take a foveated image together with all display and viewing parameters from accurate conversions of spatial frequencies. The output is an enhanced version of the foveated image. 

Our prototype in MATLAB relies on a straightforward implementation of the method for estimating the noise parameters. The noise generation is performed using the C++ implementation provided by the authors of the original paper \cite{lagae2009}. Our OpenGL implementation takes advantage of the simplicity of all the operations, which can be performed in a per-pixel fashion. The method, including the noise generation, is implemented as a series of GLSL fragment shaders. We leverage the MIP maps to efficiently implement the construction of all image pyramids used in our technique.

In all our experiments, we generate the input foveated images with the fovea region radius equal to 8 visual degrees. Beyond this region, we assumed the resolution fall-off to be linear as a function of eccentricity and defined by blur rate. The foveation was simulated using spatially-varying Gaussian blur.

Our enhancement method processes only luminance information. Consequently, the input color images are first converted to linearized luminance values which are then processed by our technique. The resulting noise contains only luminance information that is added to the input color images after gamma correction.
\section{Calibration} \label{sec:calibration}
\begin{figure}
\includegraphics[width=\linewidth]{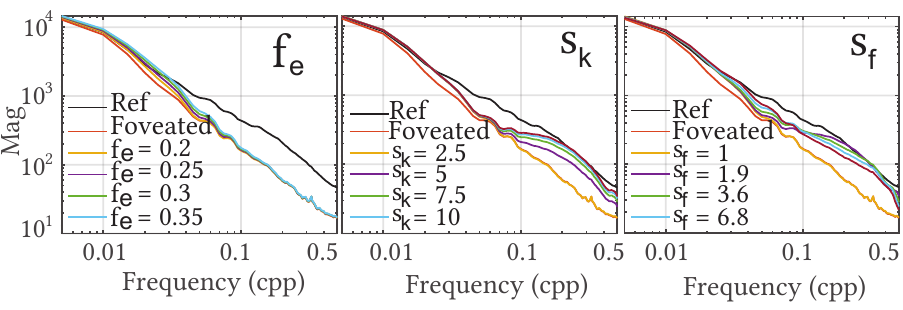}
\caption{Effect of calibration parameters on our synthesis (Sponza scene). Left: Effect of varying $f_e$ on the spatial frequency spectrum of the contrast enhanced image. Middle: Effect of varying $s_k$ on the spatial frequency spectrum of the final enhanced image for fixed $f_e$ and $s_f$. Right: Effect of varying $s_f$ on the spatial frequency spectrum of our final enhanced image for fixed $f_e$ and $s_k$.}
\label{fig:calibration_fourier}
\end{figure}

Our enhancement method contains three free parameters that allow us to fine-tune the technique. The first two are the scaling factor for the amplitude of noise, $s_k$, and the scaling factor for the noise bandwidth, $s_f$. The third parameter is the parameter of the contrast enhancement method \cite{patney2016}, $f_e$, which scales the amount of high-frequency content added to the input foveated image. While $f_e$ is fixed in the original work, the authors mention that it can be adjusted. Below, we describe the subjective experiment for estimating the values of $s_k$, $s_f$, and $f_e$. With this experiment, we aim to find the parameters such that the result of our enhancement to closely matches the full-resolution rendering. In our experiments, we assume that their optimal values solely depend on the strength of the foveation. Figure~\ref{fig:calibration_fourier} shows the effect of our calibration parameters on the spatial frequency spectrum of our enhanced image. Figure~\ref{fig:calibration_fourier} (left) shows that increasing $f_e$ increases the amount of contrast enhancement. However, as expected, the contrast enhancement has very little influence on high spatial frequencies that are lost due to foveation. Figure~\ref{fig:calibration_fourier} (middle) shows that increasing $s_k$ increases the energy of the frequencies we synthesize. Figure~\ref{fig:calibration_fourier} (right) shows that increasing $s_f$ increases the bandwidth of the distribution of frequencies we sample from and a wider range of frequencies are synthesized.

\subsection{Experiment Setup}

\paragraph{Hardware}
The experiment was conducted using LG OLED55CX. The 55-inch screen has 4K spatial and 120Hz temporal resolution. The peak luminance of the display was set to 167.33 cd/m$^2$. The position of the participant's head was fixed using a chin-rest at 71.5cm distance from the screen. This allowed covering $80\degree$ field of view and display spatial frequencies up to 24cpd at the center of the screen. We used Tobii Pro Spectrum 600Hz eye tracker to track the participants' gaze location. All the experiments were conducted under diffuse office illumination behind a dark curtain to prevent any reflections on the screen. The experimental setup is presented in Figure~\ref{fig:validation}.

\paragraph{Stimuli}
We selected five 4K images from two rendered video sequences: Big Buck Bunny\footnote{\url{https://peach.blender.org}} and Tears of Steel\footnote{\url{https://mango.blender.org}} (Figure~\ref{fig:calibration_stimuli}). We chose the images such that they include both dark and bright scenes with high-contrast and high-spatial-frequency details.

\myfigure{calibration_stimuli}{Representative images of the stimuli used in the calibration experiment.}

\paragraph{Task}
In each trial, participants' task was to adjust one of the parameters of a technique such that it best matches the reference full-resolution rendering. To this end, participants were shown half of the full-resolution image on the left. A mirrored version of the reference image was foveated using the adjusted parameter value and shown on the right. The participants were able to manipulate one of the parameters of our technique using a mouse wheel, and the image on the right (test image) was updated in real-time during the experiment to reflect the result of changes in the value. Participants were asked to adjust the parameters such that the foveated image on the right best matches the full-resolution rendering on the left, and confirm their choice by hitting the Enter key on the keyboard. During the experiment, participants were asked to keep looking at the center of the screen marked with a cross and their gaze position was constantly monitored with the eye tracker. In order to ensure the correct retinal positioning of the stimuli during the experiment, stimuli are hidden by rendering a black frame whenever participants' gaze position moved outside a small circular region with radius $1.25\degree$ at the center of the screen.

\paragraph{Participants} University students were recruited for this experiment. Prior to the experiment, they were naive about the purpose of the experiment. All of them had normal or corrected-to-normal vision, and received financial compensation for their participation. The experiment procedure was approved by the ethics committee of the hosting institution. 

\subsection{Experiment Procedure}
\label{sec:experiment_procedure}
Each procedure of adjusting a parameter is performed separately for a different amount of foveation. For adaptive-resolution foveated rendering techniques, the cutoff point of spatial frequencies shift to lower frequencies as the eccentricity increases as a result of lower resolution rendering in the periphery. In order to simulate this effect with our low pass Gaussian filter, we linearly increase the standard deviation of the Gaussian kernel with increasing eccentricity. We refer to the scalar representing this linear increase as \emph{blur-rate} in arcmin-per-degree units, and we represent it with $\sigma_\text{BlurRate} $. We define the foveal region with $ 8\degree $ radius around the gaze position where the content is rendered in full-resolution. The increase in kernel standard deviation at $8\degree$ eccentricity and beyond. We considered three levels of foveation for data collection: $\sigma_\text{BlurRate} = 0.11, 0.34, 0.57$, based on the previous analysis of Tursun\etal\shortcite{tursun2019}, which reported percentiles of foveation detectability for different blur-rates. We included the limits so that they may encompass variation of detectability across scenes. 

\begin{figure}
\includegraphics{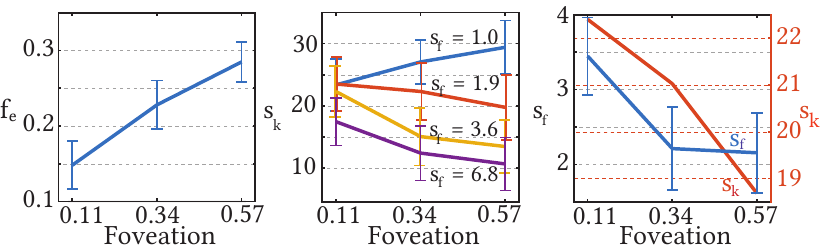}
\caption{Results of the calibration experiment. \textit{Left}: contrast enhancement parameter ($f_e$), middle: amplitude scaling ($s_k$) for different values of the noise bandwidth parameter ($s_f$), and \textit{Right}: the final combinations of amplitude scaling ($s_k$) and noise bandwidth parameter ($s_f$). The error bars visualize the standard error of the mean (SEM) for each measurement point.}
\label{fig:calibration_results}
\end{figure}

\paragraph{Contrast enhancement parameter}
We first investigated the parameter of the contrast enhancement step \cite{patney2016}, $f_e$, which scales the amount of added high-frequency signal to the foveated image. Here, we consider the contrast enhancement alone; therefore, the test image shown on the right did not contain additional noise. The participants could freely manipulate the parameter $f_e$ within the range (0,0.4). A total of five people performed the adjustment for each of the five images (Figure~\ref{fig:calibration_results}, left).

\paragraph{Amplitude scaling}
We used the estimated $f_e$ values to investigate the remaining parameters, amplitude and bandwidth scaling. Again, our design choice was to study one parameter at the time. Therefore, in the next step we asked participants to adjust $s_k$ parameter within the range (0,45) for a small set of $s_{f} = 1, 1.9, 3.61, 6.81$. The total for 10 participants adjusted the $s_k$ parameter for all combinations of five images, three $s_{f}$ values and three foveation levels (Figure~\ref{fig:calibration_results}, middle). 

\paragraph{Bandwidth scaling}
The previous experiments provided an optimal scaling factor for contrast enhancement as well as a relation between optimal amplitude, $s_k$, and bandwidth scaling, $s_{f}$, for different levels of foveation, $\sigma_\text{fov}$. To choose the final parameters for our method across different foveation levels, we ran a final experiment where participants adjust the bandwidth scaling factor, $s_{f}$, and the amplitude, $s_k$, was computed by linearly interpolating the values from the previous experiment. Again, the total of 10 participants performed this experiment (Figure~\ref{fig:calibration_results}, right).

\subsection{Results and Discussion}
Figure~\ref{fig:calibration_results} shows results obtained in the three experiments described above. The range of values obtained for the contrast enhancement parameter (left plot) matches the default parameter value, $0.2$, reported by Patney\etal\shortcite{patney2016}. The users prefer more aggressive contrast enhancement (higher $f_e$ values) as foveation increases, which can be explained by more visible loss of spatial details due to the foveation. The participants choose higher amplitude scaling ($s_k$), for our noise enhancement as the bandwidth parameter ($s_f$) decreases (middle plot). This is explained directly by the fact that the increase of the bandwidth parameter widens the spectrum of the noise beyond the aliasing range, which makes it distinguishable from the image content in the reference image. The participants compensated for this by lowering the noise amplitude. The final parameters $s_k$ and $s_f$ (right plot) demonstrate that the spectrum of the noise has to be restricted (lower $s_f$ values) more as the foveation increases. Same trend holds for corresponding amplitude scaling ($s_k$), even though the relative change in the parameter is less significant. We attribute this trend to the fact that for more aggressive foveation, the range of missing high-frequency details which are easily resolvable increases. Any attempt of compensating for this loss with noise makes it visible due to the ability of the HVS to resolve the details. Therefore, the participants perform to limit the noise-based enhancement. All the final parameters of our technique for each considered foveation level are listed in Table~\ref{tab:parameters}. A detailed statistical analysis of the user-data obtained from calibration is provided in the supplementary. 
\begin{table}[h]
    \centering
    \caption{The optimal values of our parameters after calibration.}
    \label{tab:parameters}
    \begin{tabular}{cccc}
        \toprule
        Foveation ($\sigma_\text{BlurRate} $) & $f_e$ & $s_k$ & $s_f$  \\ \midrule
        0.11 & 0.15 & 22.4 & 3.45 \\ \hline
        0.34 & 0.23 & 21.02 & 2.21 \\ \hline
        0.57 & 0.28 & 18.68 & 2.19 \\ \bottomrule
        \end{tabular}
    
\end{table}

\begin{figure*}
\includegraphics[width=\textwidth]{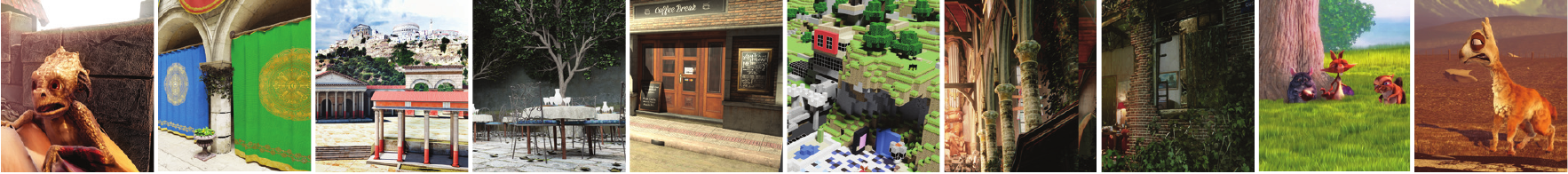}
\caption{Representative images of the stimuli used for our validation experiment.}
\label{fig:val_stim}
\end{figure*}

\begin{figure}
\includegraphics[width=1.0\columnwidth]{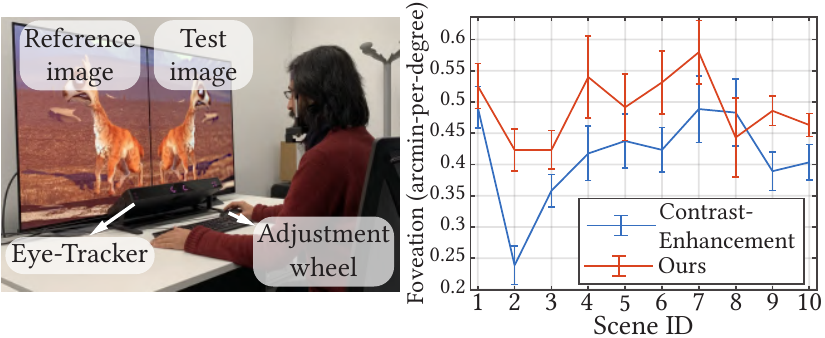}
\caption{The experimental setup of our subjective experiments is shown on the left. The average foveation strength (with higher value representing lower spatial frequency cutoff) selected by the participants in the validation experiment (Section~\ref{sec:validation_experiment}) for each scene is shown on the right. Overall, our enhancement (orange line) allows using more aggressive foveation without undesired visual artifacts compared to contrast enhancement (blue line) ($ p < 0.001 $, t-test).}
\label{fig:validation}
\end{figure}

\begin{figure*}
\includegraphics[width=\textwidth]{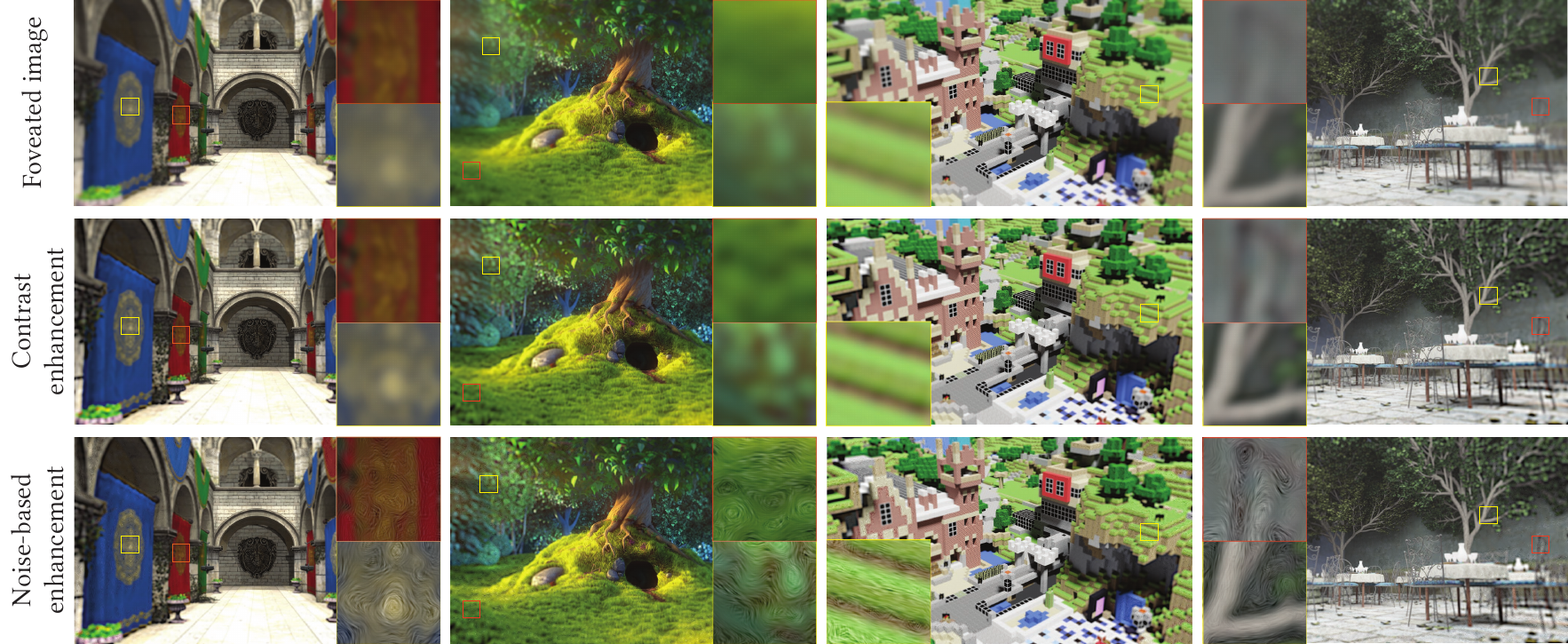}
\caption{The comparison of contrast enhancement and noise-based enhancement methods with input foveated images (generated with $\sigma_\text{BlurRate} = 0.68$). The insets show magnified views of two image regions for visual comparison. These images are designed to be viewed at an 80 degree wide field-of-view with the gaze at the center. For more results, please refer to the supplemental materials.}
\label{fig:results}
\end{figure*}

\section{Evaluation}
We evaluate the effectiveness of our noise-based enhancement with a subjective experiment. The main goal was to demonstrate that subjects can tolerate more aggressive foveation when our noise-based enhancement is used in a post-processing step. We compare our method to contrast enhancement \cite{patney2016}, which to the best of our knowledge, is the only technique with similar goals (Section~\ref{sec:related_work}), i.e., low-cost enhancement of foveated rendering. 

\subsection{Subjective Experiment}
\label{sec:validation_experiment}

\paragraph{Stimuli}
We used 10 rendered scenes to create results for contrast enhancement technique and our noise-based enhancement, which includes also contrast enhancement (Figure~\ref{fig:val_stim}). The evaluation stimuli were selected to incorporate diversity in illumination, contrast and spatial frequencies. For both techniques and for all results, we used the same parameters determined in the previous calibration experiments (Section~\ref{sec:calibration}). We simulated adaptive-resolution foveation using spatially-varying Gaussian blur whose standard deviation parameter depends on the visual eccentricity similar to Section~\ref{sec:experiment_procedure}. In Figure~\ref{fig:results}, we provide foveation ($\sigma_\text{BlurRate} = 0.45$) and enhancement results of four scenes for visual comparison (please refer to the supplemental materials for other scenes).

\paragraph{Task}
The task of each participant was to select the blur-rate at which the loss of image details due to foveation became visible with respect to the reference (full-resolution) image. To this end, in each trial we showed the original image on the left half of the display, while the right half was used for showing the output from the evaluated techniques (i.e., our method and contrast enhancement \cite{patney2016}). The stimuli were randomized during the experiment and the participants were not aware of the enhancement method being shown in any trial. The participants were asked to keep looking at a cross at the center of the screen during the experiment. The eye-tracker is used for monitoring the gaze position throughout the experiment and insuring that they adhere to the instructions (similar to our calibration experiment in Section~\ref{sec:calibration}). The participants were asked to adjust the blur-rate using the mouse-wheel (initially starting with no foveation) and press enter at the highest blur-rate at which they could not perceive foveation with respect to the reference image on the right. 

\paragraph{Hardware}
The hardware setup was identical to the equipment used in our calibration experiment (Section~\ref{sec:calibration}).

\paragraph{Participants}
10 university students, with normal or corrected-to-normal vision, were recruited for the experiment. Our test pool consisted of 3 women and 7 men. Prior to the experiment, the participants were given a written description of the experiment procedure. In addition, we explained the general concept of foveated rendering and shown a sample adaptive-resolution foveated image generated using sub-sampling and nearest-neighbor interpolation. In order to avoid introducing bias to their preferences, the sample image and the foveation method is chosen different from the techniques evaluated in the experiment. The participants received a financial compensation after completing the experiment. The whole experiment took approximately 15 minutes and completed in a single session by each participant.

\paragraph{Results}
The average values of $\sigma_\text{BlurRate}$ selected by the participants for each scene is shown on the plot in Figure~\ref{fig:validation}. The selected $\sigma_\text{BlurRate}$ values represent how ``aggressively'' high spatial frequency details can be removed from the reference full-resolution image by foveation before the loss of high frequency details become visible to the observers. In comparison to contrast enhancement, our technique allows for higher $\sigma_\text{BlurRate}$ without visible artifacts from foveation ($p<0.001$, t-test). The results of this experiment show that our method reduces the amount of perceived loss of spatial details from foveated rendering and allows for rendering in lower resolution in the periphery. For the reader's interest, we report that we did not observe any statistically significant differences based on gender.

\begin{figure}
\includegraphics[width=1.0\columnwidth]{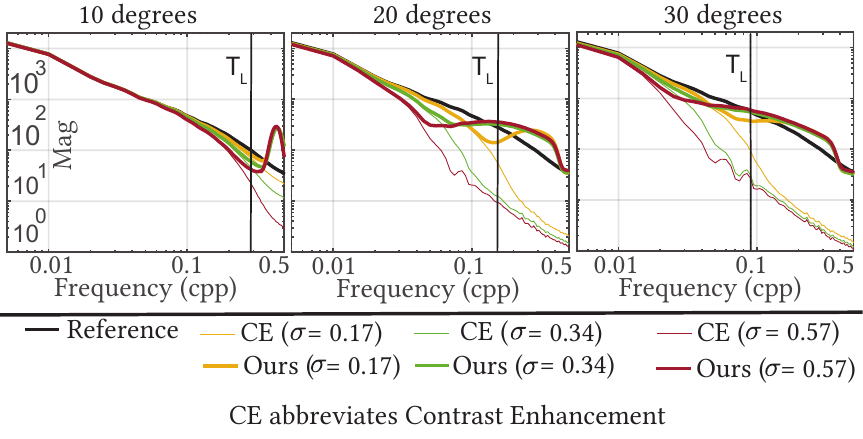}
\caption{Fourier analysis of our technique for three different foveations and eccentricities; averaged over 5 scenes. Our approach synthesizes lost high frequency content in the aliasing band which mitigates the perception of foveation.}
\label{fig:fourier_analysis}
\end{figure}

We provide an analysis of spectral characteristics of our outputs in Figure~\ref{fig:fourier_analysis}. As expected, foveation effectively removes higher-spatial frequencies which cannot be properly recovered through contrast enhancement, especially in the Aliasing Band ($f \geq T_L$). The attenuation increases with eccentricity and blur-rate, $\sigma_\text{BlurRate}$. This analysis confirms that our technique is effectively able to compensate for the lost high-frequencies. Interestingly, the noise sometimes over-compensates in the aliasing band; however, the participants did not report visual artifacts and we relate this to the frequency band being not resolvable by the HVS.

\begin{figure}
\includegraphics[width=\linewidth]{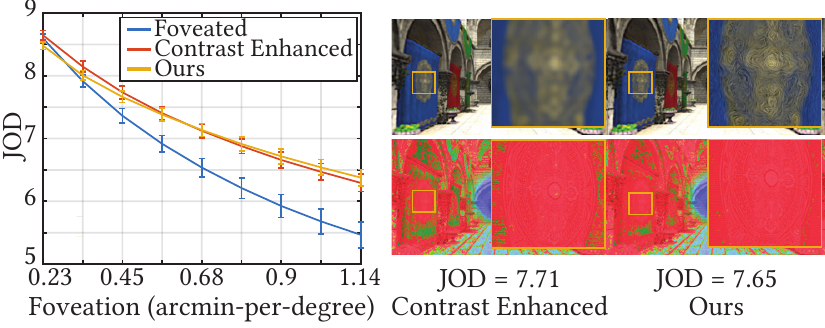}
\caption{Output of FovVDP for our noise-based enhancement, contrast enhancement, and standard foveated rendering results (FovVDP reference: full-resolution image). \textit{Left}: FovVDP produces very similar JOD scores for our method and contrast enhancement, whereas standard foveation always has worse JOD scores. \textit{Right}: FovVDP input images and error maps for contrast enhancement and our method with similar JOD scores at the same amount of foveation ($\sigma_\text{BlurRate} = 0.68$).}
\label{fig:fovvdp_outputs}
\end{figure}

Additionally, we report the outputs from FovVDP \cite{mantiuk2021} objective quality metric for our method, contrast enhancement, and standard foveation in Figure~\ref{fig:fovvdp_outputs} (FovVDP reference: full-resolution image). Overall, the average JOD scores obtained from FovVDP are significantly different between the standard deviation and both contrast enhancement and noise enhancement methods ($p<0.001$, t-test). However, visual quality difference between the contrast enhanced and noise-enhanced images are not significantly different ($p=0.07$, t-test). We believe that relatively close JOD scores obtained for our method and contrast enhancement is due to the greatly reduced contrast sensitivity function used in FovVDP for the frequencies in the aliasing band. We find this observation intuitive because contrast sensitivity functions usually represent the measurements from psychovisual experiments where the visibility threshold is defined by resolvable contrast rather than detectable contrast. Hence, we believe that FovVDP and similar objective quality metrics are not suitable references for evaluating the visual quality of methods that enhance the aliasing band. This is further supported by the results of our subjective evaluation which indicates difference between our and contrast enhancement method.

\subsection{Temporal Coherence}
During the derivation of our method, we took various steps to mitigate potential issues regarding temporal stability. Our method for estimating the noise parameters relies on simple image processing operations, e.g., Gaussian/Laplacian pyramid and Sobel filter. Due to their filtering nature, these operations usually preserve the temporal coherence of the input image sequence and generate stable results as long as the input is stable. Furthermore, the frequency spectrum of our noise is defined using a smooth log-normal distribution which prevents any abrupt changes. We also fixed the impulses locations for the noise generation for the duration of a rendering sequence. As a result, the locations of the Gabor patches do not change, and only their parameters adapt to the underlying content. The static nature of our impulse locations could hint towards motion artifacts such as the shower-door effect, which occurs when static content over-layed on videos disturbs motion cues. However, even-though the impulse locations are static over frames, the per-impulse noise parameters such as orientation and amplitude change smoothly with the underlying motion. Therefore, our noise pattern is not static but dynamically adjusting according to the underlying content, mitigating the shower-door effect. Finally, our noise is generated for less sensitive regions of our visual field, which makes potential temporal problems less visible. However, we want to clarify that our technique does not explicitly aim at exact reproduction of motion cues present in full-res videos, which in fact is a much more complex open-problem and an exciting venue for future work (Section 7).

\begin{figure}
\includegraphics[width=\linewidth]{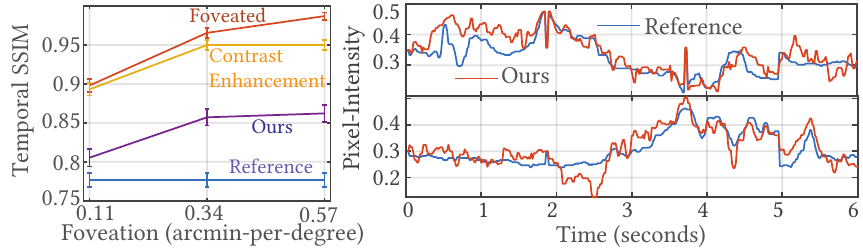}
\caption{The figure demonstrates the temporal coherence of our method compared to full-resolution rendering, foveated input, and contrast enhancement. On the right, we demonstrate temporal slices of pixel intensities for two image locations. On the left, we measure the temporal coherence by computing SSIM metric values between consecutive frames and averaging them across the image sequence.}
\label{fig:temporal_coherence}
\end{figure}

To the best of our knowledge, there is no established metric for evaluating the temporal stability of a video sequence, especially for wide-field-of-view setups. To hint at the temporal coherency of our solution, we first visualize the temporal pixel intensity variation for two locations in one of our sequences and compare it to the corresponding signal in the full-resolution rendering. Figure~\ref{fig:temporal_coherence} (right) shows the results of this analysis. While the temporal variation for our technique is slightly larger than in the full-resolution rendering, the deviation remains low without creating significant temporal instabilities. 

Additionally, we follow the procedure used in previous work \cite{kaplanyan2019}, where the temporal consistency was measured using SSIM metric \cite{Wang2004} values between consecutive frames averaged across an image sequence. Figure~\ref{fig:temporal_coherence} (left) shows the results of this analysis for one sequence and different levels of foveations. For comparison, we report values for our method, full-resolution rendering, foveated input, and when only the contrast enhancement is applied. The foveated sequence has the highest average SSIM between frames because of spatial smoothing introduced by foveation. It is followed by the contrast enhancement, our method and ground truth, respectively. As our method synthesizes noise over the contrast enhanced frames, a decrease in SSIM with respect to the contrast enhancement is expected. Still, our method delivers high average inter-frame SSIM when compared to the reference sequence. 

The above analysis gives only hints on the temporal properties of our technique. We refer to our supplemental materials for rendering sequences processed with our method.

\subsection{Performance}
\setlength{\columnsep}{10pt}
\begin{wrapfigure}{r}{0.0pt}
  \centering
    \includegraphics{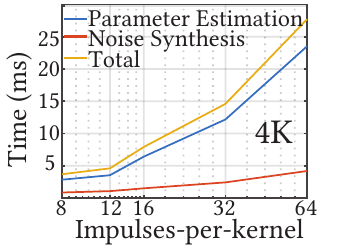}
  \caption{Running times}
  \label{fig:performance}
\end{wrapfigure}
We evaluated the performance of the noise-based enhancement method using a PC machine with Intel(R) Core(TM) i7-9700K CPU at 3.60GHz, 32 GB of RAM, and a single NVIDIA GeForce RTX 2080 Ti GPU. Considering the practical requirements of foveated rendering, we measured the performance for 4K resolution (3840$\times$2160\,px) input images. We considered the two main steps of our method: noise parameter estimation and noise synthesis. The parameter estimation component consists of building all the image pyramids and estimating frequency, amplitude, and orientation for each Gabor patch. The synthesis considers sparse convolution of impulses with Gabor patches. Furthermore, we consider different impulse densities for the noise generation as it provides a trade-off between the noise quality and performance (Section~\ref{sec:noise_generation}). The corresponding performance plots are shown in Figure~\ref{fig:performance}. We can observe that the main cost of our technique is the parameter estimation for the noise. Overall, the performance also increases roughly linearly as the number of kernels increases. Note the running times are plotted in a logarithmic domain. Our initial choice of 64 impulses per kernel was motivated by the recommendation in the original paper \cite{lagae2009}. However, this recommendation did not take into account the possibility of showing the noise in the periphery. Consequently, we experimented with lowering the number of impulses. Our initial visual inspection revealed that the noise pattern preserves its quality even for as little as 12 impulses per kernel (Figure\ref{fig:kernels_density}, left). To evaluate that this number is still sufficient for our noise-based enhancement technique, we repeated the validation experiment (Section~\ref{sec:validation_experiment}) for 12 impulses. The results of this experiment (Figure\ref{fig:kernels_density}, right) demonstrate that the advantage provided by our method over the contrast-enhancement method remains unchanged when compared to the experiment for 64 impulses ($p<0.001$, t-test). This suggests that reducing the number of impulses to 12 is a viable option for improving the performance of our method. This brings down the time required for running the entire enhancement process to $4.6ms$ (217 frames per second), compared to $27.7ms$ (36 frames per second) for 64 impulses.

The run-time evaluation presented here was performed for a straightforward implementation of our method on GPU. We believe that further gains can be obtained when the code is optimized. We report exact scene-independent execution times of our method. The actual gains in the foveated rendering pipeline depend significantly on factors such as scene-complexity, geometry-cost, and the complexity of shading (e.g volumetric scattering). While for some very simple scenes, foveated rendering gains may be very low, scenes requiring expensive shading are likely to deliver higher gains [Tursun et al. 2019]. For ray-tracing however, it is possible to approximately correlate our reported blur-rate improvement with the sampling-rate reduction. Assuming a spatial gaussian cut-off of 2$\cdot\sigma(x)$ at location $x$, the corresponding sampling-rate according to the Nyquist-rate would be

\begin{equation}
  SR(\mathbf{x}) = \frac{1}{4\cdot\sigma(\mathbf{x})}
\end{equation}

After integration over the image, we can conclude that the net sampling-rate ($\hat{SR}$) reduction ratio is the same as the blur-rate ($\sigma_{BlurRate}$) increase ratio i.e 

\begin{equation}
  \frac{\hat{SR}_2}{\hat{SR}_1} = \frac{\sigma_{BlurRate}^1}{\sigma_{BlurRate}^2}
\end{equation}

Considering Eq. 11, we report estimated reduction in sampling-rate's for our technique over contrast enhancement in Figure ~\ref{fig:kernels_density}-right. 

It is also important to note that for VR-HMDs, which require stereo rendering, our noise synthesis is performed only once followed by inexpensive warping [Didyk et al. 2011]. Furthermore, we believe that future higher resolution (e.g 8K) displays will benefit even more from our technique. 

Our technique is designed for the real-time foveated rendering pipeline. However, it may also find application in foveated compression. Regarding bandwidth savings of our technique, we report up-to $23\%$ reduction in foveated image file-size over contrast enhancement for the images used in our evaluation. The blur-rates used were in accordance with user-experiments in Figure 16-right. The saving was estimated by compressing foveated images using JPEG at a quality factor of 90 and comparing the file sizes.

\begin{figure}
\includegraphics[width=\linewidth]{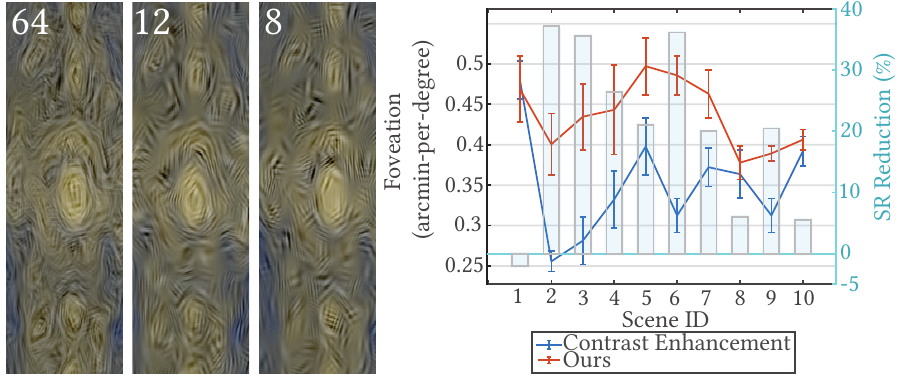}
\caption{Reducing the number of impulses per kernel for noise generation has little effect on the quality (left). The results of the additional experiment (right) demonstrate that even enhancement based on 12 impulses per kernel allows more aggressive foveation, and the gains are similar to those obtained for 64 impulses ($p<0.001$, t-test). The bar plot shows the estimated sampling-rate reduction of our technique over contrast enhancement.}
\label{fig:kernels_density}
\end{figure}

\subsection{Stereoscopic Content}
\begin{figure}
\includegraphics[width=\linewidth]{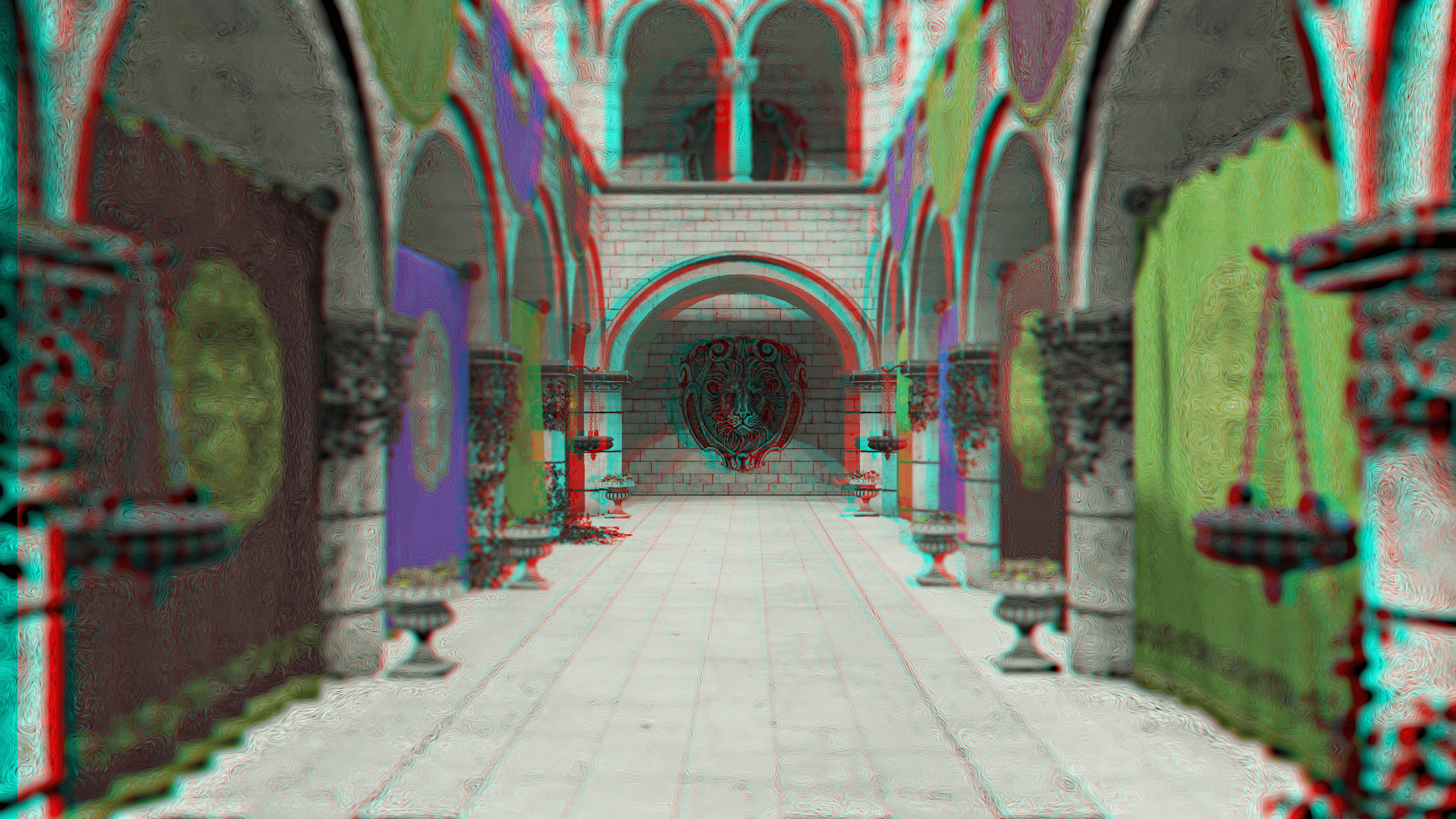}
\caption{The image shows our technique applied to a stereoscopic rendering of the Sponza scene. The image is presented in red-cyan anaglyph colors. This image is designed to be viewed at an 80 degree wide field-of-view, with gaze at the center.}
\label{fig:stereo_fig}
\end{figure}

Application to virtual and augmented reality devices requires stereoscopic rendering. The results presented so far consist of monoscopic images. As a simple extension to our technique for stereoscopic rendering, we propose to generate the noise only for one eye and use an image-based warping technique \cite{Didyk2010} to generate the noise for the second eye by warping the noise according to the disparity. Such a technique creates an illusion of noise being a part of the object surfaces and textures. Due to the random nature of our noise, we did not observe visible problems at disocclusion regions. Figure ~\ref{fig:stereo_fig} shows a red-cyan anaglyph version of our enhancement applied on a stereoscopic image. Please refer to supplemental materials for more examples of our technique applied to stereoscopic content.

\section{Limitations and Future Work}
\label{sec:limitations}

Our technique estimates the parameters of the noise based on the available information in a foveated image. In particular, it propagates the statistics such as spatial frequency amplitude and orientation from lower frequencies to higher ones. This process relies on the assumption that low and high spatial frequencies share similar statistics. However, this does not always need to be the case. Some parts of the scene or textures may inherently contain only low-frequency content, for example, sky, water, or image regions with a depth-of-field effect. Despite this, our technique will still generate high-spatial frequency noise. Scene 10 in our evaluation validation experiment (Section~\ref{sec:validation_experiment}) is a particular example of it. In future work, it would be interesting to look for ways to address this limitation. For example, the particular case of the depth-of-field effect could be addressed by exploiting the camera parameters used for rendering the scene during the estimation of the noise parameters. Similarly, the information about frequency spectra of textures could be precomputed and stored prior to the rendering. This would inform the noise-based enhancement technique not to generate noise for regions containing only low-frequency content.

\begin{figure}
\includegraphics[width=\linewidth]{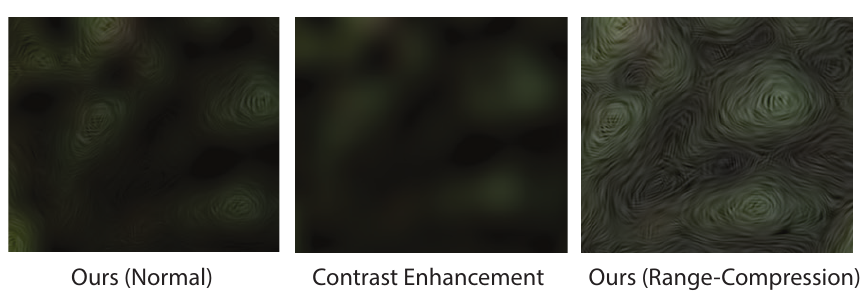}
\caption{Saturation $\&$ Noise ($Scene$-8). \textit{Left}: Our enhancement after contrast enhancement leaves less range for the noise to be applied.\textit{Center}:  Saturation after contrast enhancement. \textit{Right}: Dynamic range reduction after contrast enhancement leaves more range for noise.}
\label{fig:clipping_problem}
\end{figure}

The efficiency of our enhancement method is limited in very high-contrast regions where adding noise leads to clipping image values. Our method accounts for such situations by attenuating the noise amplitude. However, such a strategy may prevent enhancement in very dark or bright image regions. Examples of such cases can be found in Scenes 1 and 8 in our validation experiment. The scenes contain dark regions that are clipped after contrast enhancement, which prevents adding noise. It would be interesting to investigate the possible trade-off between the dynamic range and our enhancement. In particular, it is possible to compress local image contrast such that the noise can be added. We provide an example of such an operation in Figure~\ref{fig:clipping_problem}.

Our technique assumes that the input is a foveated rendering sequence that does not contain any spatial or temporal aliasing. Furthermore, following the studies of Albert\etal\shortcite{albert2017}, Hoffman\etal \shortcite{hoffman2018} and the previous work on foveated rendering \cite{patney2016,tursun2019}, we simulate and model foveated rendering using spatially-varying Gaussian blur. An exciting venue for future work is to investigate the enhancement of the foveated rendering containing some form of aliasing. An analysis of the aliasing could provide additional information to the noise generation technique on the range of spatial frequencies that were not correctly synthesized. 

Most of the widely available virtual reality headsets provide limited spatial resolution, especially for large eccentricity \cite{Beams2020}. Therefore, we performed all experiments using a large 4K screen, which, for our viewing conditions, can reproduce spatial frequencies up to 24 cycles per visual degree for the fovea region. In comparison, the upper limit on the spatial frequency reproduction for the HTC Vive Pro Eye headset is 5 cycles per visual degree, which is insufficient to fully exploit the benefits of our enhancement. Newer display models such as the Varjo VR-3 HMD can represent up to 15 cycles-per-degree in the periphery, which is sufficient to exploit the full benefits of our method.

In this work, we demonstrate the utility of our technique in the context of rendering. However, we believe that in a similar way, the method can be used to enhance foveated video compression and complement such techniques as \cite{walton2021}. Furthermore, we believe that there is significant potential in the application of noise-based techniques for many open problems in real-time foveated graphics. We have only yet explored temporally-coherent spatial characteristics of Gabor-noise for our technique, whereas variations such as 3D Gabor-noise can effectively control temporal characteristics of the content. More specifically, there is often a trade-off between ensuring temporal coherency and preserving motion-cues. One example is temporal anti-aliasing which provides temporal coherency in rendered videos by removing temporal artifacts at the cost of damping some motion cues. Exact reproduction of motion-cues present in full-resolution renderings i.e temporal metamers, is still an open-problem, but an exciting potential extension of our work. 

Considering the spectral nature of various psycho-visual models, the ability to precisely control spatio-temporal spectral properties of synthesized content presents an exciting tool to bridge human perception and graphics.

\section{Conclusions}
Foveated rendering enables significant computation benefits for high-resolution wide-field-of-view displays. Therefore, it becomes a crucial component and key enabler for high-quality rendering systems in virtual and augmented reality headsets. In this work, we presented an efficient technique that can further reduce the cost of the foveated rendering by allowing more aggressive foveation without perceived quality loss. We achieve this goal thanks to our new noise-based enhancement step, which replaces rendering of high-spatial frequency details with inexpensive procedural noise. This paper presents a perceptually-inspired method for estimating the parameters of the noise and calibration of the technique based on data collected in perceptual experiments. The method consists of a series of simple image processing steps applied directly to the output of the foveated rendering. Thanks to these properties, our enhancement method is suitable for direct integration into foveated rendering systems.

\appendix
\section{Detection and Resolution Frequency Threshold Measurements}
\label{sec:thibos_measurements}

\begin{table}[h!]
    \centering
    \caption{The measurements from Thibos\etal\shortcite{thibos1996}, that we linearly interpolate to obtain $T_H(e_\mathbf{x})$ and $T_L(e_\mathbf{x})$ for different values of $e_\mathbf{x}$.}
    \label{tbl:thibos_values}
    \begin{tabular}{r|ccccccc}
    \toprule
    Eccentricity (deg) & 0  & 5  & 10   & 15 & 20  & 25  & 30   \\ \hline
    $T_L$              & 60 & 27 & 10.5 & 8  & 5.5 & 4.8 & 4    \\ \hline
    $T_H$              & 60 & 40 & 26   & 24 & 23  & 21  & 20.5 \\ \bottomrule
    \end{tabular}
\end{table}

\bibliographystyle{ACM-Reference-Format}


\end{document}